\documentclass{article}

\PassOptionsToPackage{numbers, compress}{natbib}

\usepackage[dblblindworkshop, final]{neurips_2025}
\workshoptitle{Workshop on AI and ML for Next-Generation Wireless Communications and Networking}

\usepackage[utf8]{inputenc} 
\usepackage[T1]{fontenc}    
\usepackage{hyperref}       
\usepackage{url}            
\usepackage{booktabs}       
\usepackage{amsfonts}       
\usepackage{nicefrac}       
\usepackage{microtype}      
\usepackage{xcolor}         

\usepackage{booktabs,tabularx,adjustbox, amssymb}
\newcolumntype{L}[1]{>{\raggedright\arraybackslash}p{#1}}
\newcolumntype{C}[1]{>{\centering\arraybackslash}p{#1}}
\usepackage{amsmath,amsfonts}
\usepackage{algpseudocode}
\usepackage{algorithm}
\usepackage{array}
\usepackage{balance}
\usepackage{cleveref}
\usepackage{makecell}
\usepackage{multirow}
\usepackage[caption=false,font=normalsize,labelfont=sf,textfont=sf]{subfig}
\usepackage{textcomp}
\usepackage{stfloats}
\usepackage{url}
\usepackage{verbatim}
\usepackage{graphicx}
\usepackage{booktabs, makecell} 
\usepackage[acronyms,toc]{glossaries}
\usepackage{soul}  
\usepackage{verbatim}
\usepackage[pdf]{pstricks}
\usepackage{comment}

\title{From Simulation to Practice: Generalizable Deep Reinforcement Learning for Cellular Schedulers}

\author{%
  Petteri Kela \\
  Nokia, Espoo, Finland \\
  \texttt{petteri.kela@nokia.com}
  \And
  Bryan Liu \\
  Nokia Bell-Labs, Massy, France \\
  \texttt{bryan.liu@nokia-bell-labs.com}
  \And
  Alvaro Valcarce \\
  Nokia Bell-Labs, Massy, France \\
  \texttt{alvaro.valcarce\_rial@nokia-bell-labs.com}
}

\begin{document}

\maketitle

\newacronym{1L}{1LDS}{Single Loop Deep Scheduler}
\newacronym{2L}{2LDS}{Double Loop Deep Scheduler}
\newacronym{3GPP}{3GPP}{3rd Generation Partnership Project}
\newacronym{5G}{5G}{Fifth Generation}
\newacronym{5GNR}{5G NR}{5G New Radio}
\newacronym{5GQI}{5GQI}{5G QoS Indicator}
\newacronym{AB}{AB}{Action Branching}
\newacronym{AC}{AC}{Actor-Critic}
\newacronym{ACK}{ACK}{Acknowledgement}
\newacronym{AI}{AI}{Artificial Intelligence}
\newacronym{BLER}{BLER}{Block Error Rate}
\newacronym{BT}{BT}{Bursty Traffic}
\newacronym{BTS}{BTS}{Base Transceiver Station}
\newacronym{C-RNTI}{C-RNTI}{Cell Radio Network Temporary Identifier}
\newacronym{CE}{CE}{Control Element}
\newacronym{CSI}{CSI}{Channel State Information}
\newacronym{CQI}{CQI}{Channel Quality Indicator}
\newacronym{DQN}{DQN}{Deep Q-Network}
\newacronym{DDQN}{DDQN}{Double Deep Q-Network}
\newacronym{DCI}{DCI}{Downlink Control Information}
\newacronym{DDPG}{DDPG}{DDPG}
\newacronym{DL}{DL}{Downlink}
\newacronym{DL-SCH}{DL-SCH}{Downlink Shared Channel}
\newacronym{DRL}{DRL}{Deep Reinforcement Learning}
\newacronym{DSACD}{DSACD}{Distributional Soft Actor-Critic Discrete}
\newacronym{DRX}{DRX}{Discontinuous Reception}
\newacronym{ERM}{ERM}{Experience Replay Memory}
\newacronym{FB}{FB}{Full Buffer}
\newacronym{FC}{FC}{Fully Connected}
\newacronym{FD}{FD}{Frequency Domain}
\newacronym{FDS}{FDS}{Frequency Domain Scheduling}
\newacronym{FR1}{FR1}{Frequency Range 1}
\newacronym{FTP}{FTP}{File Transfer Protocol}
\newacronym{GAE}{GAE}{Generalized Advantage Estimator}
\newacronym{GBR}{GBR}{Guaranteed Bit Rate}
\newacronym{gNB}{gNB}{gNodeB}
\newacronym{GoB}{GoB}{Grid of Beams}
\newacronym{HARQ}{HARQ}{Hybrid ARQ}
\newacronym{JSD}{JSD}{Jensen-Shannon Divergence}
\newacronym{KNN}{KNN}{K-Nearest Neighbors}
\newacronym{KPI}{KPI}{Key Performance Indicator}
\newacronym{L2C}{L2C}{Learning to Communicate}
\newacronym{LCID}{LCID}{Logical Channel ID}
\newacronym{LTE}{LTE}{Long Term Evolution}
\newacronym{MAC}{MAC}{Medium Access Layer}
\newacronym{MARL}{MARL}{Multi-Agent Reinforcement Learning}
\newacronym{MCS}{MCS}{Modulation and Coding Scheme}
\newacronym{MCTS}{MCTS}{Monte Carlo Tree Search}
\newacronym{MIMO}{MIMO}{Multiple Input Multiple Output}
\newacronym{ML}{ML}{Machine Learning}
\newacronym{MNO}{MNO}{Mobile Network Operator}
\newacronym{MRC}{MRC}{Maximal-ratio combining}
\newacronym{MSE}{MSE}{Mean Square Error}
\newacronym{MUMIMO}{MU-MIMO}{Multi-User MIMO}
\newacronym{PDB}{PDB}{Packet Delay Budget}
\newacronym{PDCCH}{PDCCH}{Physical Downlink Control Channel}
\newacronym{PDCP}{PDCP}{Packet Data Convergence Protocol}
\newacronym{PDSCH}{PDSCH}{Physical Downlink Shared Channel}
\newacronym{PDU}{PDU}{Protocol Data Unit}
\newacronym{PF}{PF}{Proportional Fair}
\newacronym{PG}{PG}{Policy Gradient}
\newacronym{PHY}{PHY}{Physical Layer}
\newacronym{PMI}{PMI}{Precoder Matrix Indicator}
\newacronym{PPO}{PPO}{Proximal Policy Optimization}
\newacronym{PUCCH}{PUCCH}{Physical Uplink Control Channel}
\newacronym{PUSCH}{PUSCH}{Physical Uplink Shared Channel}
\newacronym{QoS}{QoS}{Quality of Service}
\newacronym{RACH}{RACH}{Random Access Channel}
\newacronym{RAR}{RAR}{Random Access Response}
\newacronym{RB}{RB}{Resource Block}
\newacronym{RBG}{RBG}{Resource Block Group}
\newacronym{RI}{RI}{Rank Indicator}
\newacronym{RL}{RL}{Reinforcement Learning}
\newacronym{RLC}{RLC}{Radio Link Control}
\newacronym{RR}{RR}{Round Robin}
\newacronym{RRC}{RRC}{Radio Resource Control}
\newacronym{RSRP}{RSRP}{Reference Signal Received Power}
\newacronym{RZF}{RZF}{Regularized Zero Forcing}
\newacronym{SAC}{SAC}{Soft Actor-Critic}
\newacronym{SACD}{SACD}{Soft Actor-Critic Discrete}
\newacronym{SAW}{SAW}{Stop-And-Wait}
\newacronym{SD}{SD}{Spatial Domain}
\newacronym{SDS}{SDS}{Spatial Domain Scheduling}
\newacronym{SE}{SE}{Spectral Efficiency}
\newacronym{SL}{SL}{Single Loop}
\newacronym{SoTA}{SoTA}{State-of-the-Art}
\newacronym{SSSG}{SSSG}{Search Space Set Group}
\newacronym{SDU}{SDU}{Service Data Unit}
\newacronym{SINR}{SINR}{Signal-to-Interference-and-Noise Ratio}
\newacronym{SLS}{SLS}{System-Level Simulator}
\newacronym{SNR}{SNR}{Signal-to-noise Ratio}
\newacronym{SS}{SS}{Synchronization Signal}
\newacronym{SUMIMO}{SU-MIMO}{Single-User MIMO}
\newacronym{TB}{TB}{Transport Block}
\newacronym{TD}{TD}{Time Domain}
\newacronym{TBS}{TBS}{Transport Block Size}
\newacronym{TDD}{TDD}{Time Division Duplexing}
\newacronym{TDS}{TDS}{Time Domain Scheduling}
\newacronym{TS}{TS}{Technical Specification}
\newacronym{TTI}{TTI}{Transmission Time Interval}
\newacronym{UCI}{UCI}{Uplink Control Information}
\newacronym{UE}{UE}{User Equipment}
\newacronym{UPT}{UPT}{User Perceived Throughput}
\newacronym{UL}{UL}{Uplink}
\newacronym{UL-SCH}{UL-SCH}{Uplink Shared Channel}
\newacronym{XR}{XR}{Extended Reality}
\newacronym{VR}{VR}{Virtual Reality}
\newacronym{vRAN}{vRAN}{virtualized Radio Access Network}
\newacronym{ZF}{ZF}{Zero Forcing}

\begin{abstract}
Efficient radio packet scheduling remains one of the most challenging tasks in cellular networks, and while heuristic methods exist, practical deep learning–based schedulers that are 3GPP-compliant and capable of real-time operation in 5G and beyond are still missing.
To address this, we first take a critical look at previous deep scheduler efforts.
Secondly, we enhance \gls{SoTA} deep \gls{RL} algorithms and adapt them to train our deep scheduler. In particular, we propose a novel combination of training techniques for \gls{PPO} and a new \gls{DSACD} algorithm, which outperformed other variants tested.
These improvements were achieved while maintaining minimal actor network complexity, making them suitable for real-time computing environments.
Furthermore, entropy learning in SACD was fine-tuned to accommodate resource allocation action spaces of varying sizes.
Our proposed deep schedulers exhibited strong generalization across different bandwidths, number of \gls{MUMIMO} layers, and traffic models.
Ultimately, we show that our pre-trained deep schedulers outperform their heuristic rivals in realistic and standard-compliant 5G system-level simulations.
\end{abstract}

\section{Introduction}

Wireless systems have evolved and are capable of serving vast amounts of data to many users over large areas. This progress stems from splitting wireless channels into resources across multiple dimensions: code, frequency, time, antennas, beams, and spatial layers. In 4G, 5G, and, likely, 6G, the radio packet scheduler dynamically allocates these resources to users, optimizing network performance, fairness, and spectral efficiency.
We refer to these functions as \gls{TDS}, \gls{FDS}, and \gls{SDS}.
They are NP-hard combinatorial problems, often solved with heuristics (see \cite{4556220}) that lack optimality guarantees and scale poorly. 
As we transition to 6G, the number of radio resources at our disposal will continue to grow (e.g., larger \gls{MIMO} arrays, larger bandwidths, diverse user devices, etc.).
This increasing complexity strains traditional scheduling methods, creating a compelling need for new approaches.
\gls{ML} offers a promising path forward for several reasons:
\begin{itemize}
    \item \textbf{Scalability}: \gls{ML} shifts the computational cost away from inference. Trained models can often make complex scheduling decisions cheaper.
    \item \textbf{Flexibility}: \gls{ML} models naturally consider multimodal factors, such as \gls{CSI}, \gls{UE} priorities, \gls{QoS} profiles, etc.
    \item \textbf{Adaptability and customization}: \gls{ML} models learn to adapt to changing channel conditions, traffic patterns, and user demands, a critical capability as networks become more dynamic.
\end{itemize}
The ability of packet schedulers to react quickly to load changes while considering all other inputs is hard to balance and build into heuristic algorithms.
For instance, some radio schedulers may select the best beam based on beam-specific \gls{CSI} reports.
Grounding this decision also on the traffic load per beam (even if the beam has worse RF conditions) can sometimes yield net capacity gains \cite{ramin_beam_management}.
Such load changes can happen very fast (especially in the higher FR2 frequencies at 24.25 GHz to 52.6 GHz) when \glspl{UE} move across beams very quickly.
Detecting such rapid changes requires additional subroutines built into heuristic schedulers, which increase their computational complexity and make debugging even harder.
Instead, \gls{ML}-based deep schedulers trained on real scenarios excel at detecting such hidden patterns in the data, often before the event occurs.
While beam selection is a crucial aspect of scheduling in \gls{GoB} systems, our current paper addresses the broader challenge of resource allocation in a dynamic \gls{ZF}-based \gls{MUMIMO} beamforming context, particularly well suited for \gls{FR1} (below 6 GHz) and the emerging 6G mid-bands (7-15 GHz).
In the remainder of this paper, we will refer to \gls{ML}-based radio packet schedulers as \emph{Deep Schedulers}.
We provide an overview of current \gls{SoTA} heuristic schedulers and previous work on deep schedulers.
We then detail our own \gls{RL}-based deep scheduler designs, focusing on schedulers based on \gls{SACD} and \gls{PPO}, which achieve competitive \gls{MUMIMO} gains while maintaining manageable computational complexity.
Finally, we compare various approaches and discuss the simulation results obtained with a \gls{5GNR}-compliant \gls{SLS}.

\section{Background}

To contextualize our work, we first review existing scheduling approaches (see \Cref{DS_history}, Appendix \ref{sec:ds_history}, for a summary).
Our focus is on developing deep schedulers suitable for practical deployment in \gls{3GPP}-compliant cellular networks (\gls{5G} and beyond), as non-standard approaches have limited real-world applicability. 
The following review highlights key limitations of the prior art concerning complexity, scalability, standard adherence, and performance, motivating the need for the advancements presented in this paper.

\subsection{Previous Deep Scheduler Efforts}    
The field of \gls{ML}-based radio resource allocation has a rich history, with numerous designs proposed over the years \cite{10293754}.
However, the complexity of wireless scheduling makes optimal solutions computationally challenging, hindering the acquisition of reliable labeled data for supervised learning models.
The time-sequential nature of scheduling decisions lends itself to hidden Markov chains, leading most deep schedulers to employ \gls{RL}.

In \cite{9110842}, a \gls{DL} \gls{FD} deep scheduler based on \gls{DDQN} was proposed, trained to maximize a composite reward for bitrate and fairness via non-contiguous allocation of 5G-compliant \glspl{RBG}. 
However, it ignored subband-specific \glspl{CQI}, missing frequency selectivity opportunities.
Its Q network had two \gls{FC} hidden layers of size 128 with ReLU activations, performing comparably to a \gls{PF} baseline.
The paper did not discuss inference time or computational complexity, which is crucial for 5G schedulers making decisions every slot with sub-ms latency.

The deep scheduler proposed in \cite{10247079} addresses \gls{SDS} in \gls{UL} \gls{MUMIMO} settings using \gls{SAC} and \gls{KNN}, with the aim of maximizing \gls{SE} and fairness among users.
It incorporates \gls{UE} grouping based on channel correlations and is evaluated with simulated and real-world data.
However, the proposed method deviates from the \gls{5GNR} uplink allocation requirements, which mandate contiguous or almost contiguous \gls{RB} allocations with \gls{RBG} granularity \cite{3gpptr38101}.
This discrepancy in the uplink allocation strategy limits the direct applicability of the approach within standard-compliant \gls{5GNR} systems.

Other efforts to reduce the action space size of deep schedulers in \gls{MUMIMO} settings include \glspl{DQN} with \gls{AB} \cite{Ericsson2024}.
However, this method does not scale effectively to practical massive MIMO systems, as it requires a large action space even with just two MIMO layers, involving $N_a$ actions and 3.6 million trainable parameters.
Additionally, \gls{RL}-based methods incorporating \gls{MCTS} have also been explored \cite{9838584}.

More recently, a hybrid \gls{FD} deep scheduler was proposed \cite{QoSdrama}, where the \glspl{UE} are selected heuristically according to a weighted priority metric.
A \gls{PPO} agent is trained to learn the heuristic's priority weights to minimize delay and packet loss.
The PPO agent used five hidden layers of size 64.
By leveraging the priority-based heuristic, the priority weights can change slowly and the inference pass does not need to be executed in each slot.
Such an approach overcomes most implementation challenges, but sacrifices some dynamicity and may lead to suboptimal decisions due to the inherent limitations of the priority-based heuristic.

Recognizing the large combinatorial action space that radio schedulers confront, the authors in \cite{10329922} have also opted for a policy gradient method such as \gls{DDPG}, which uses an \gls{AC} architecture.
The aim is to train a \gls{ML} model for aiding the scheduler to estimate an adequate allocation size $N_{\text{RBs}}$ and \gls{MCS} for each \gls{UE} under \gls{vRAN} computational constraints. This design considers the \gls{UE}-specific uplink \gls{SNR} and the congestion of the cloud platform's CPUs, and it rewards the model for maximizing successful data decoding, thus promoting high bitrate.
The authors claim significant gains in spectral efficiency over a \gls{RR} baseline.

Despite these efforts, none are practical enough for deployment in commercial cellular networks. 
We argue that a practical Deep Scheduler should possess the following traits:
\begin{itemize}
    \item \textbf{Sub-ms latency}: Light inference passes.
    \item \textbf{Diverse traffic performance}: Effective handling of bursty and \gls{FB} traffic.
    \item \textbf{\gls{SDS}}: Support for \gls{MUMIMO} in massive MIMO systems.
    \item \textbf{Rank-aware scheduling}: Maintains performance across channels with varying diversity.
    \item \textbf{Frequency-selective scheduling}: In a standards-compliant manner.
\end{itemize}
Our current work focuses on \gls{FDS} and \gls{SDS} challenges. Extending similar \gls{ML}-based principles to \gls{TDS} functions could enable a fully integrated Deep Scheduler, including:
\begin{itemize}
    \item Logical channel prioritization.
    \item \gls{PDCCH} \& \gls{PDSCH} co-scheduling.
    \item QoS-driven scheduling: Effective handling of \gls{GBR} and non-\gls{GBR} traffic, supporting standard QoS profiles.
\end{itemize}

\section{Practical Deep Scheduler Implementation and Analysis}

We train our ML-based schedulers with on-policy and off-policy \gls{RL} algorithms combining several recent advances in \gls{RL} research.
We have evaluated two different designs for \gls{RL}-based scheduling. Either looping through all \glspl{RBG} and \gls{MUMIMO} user layers or increasing our input and output layers to accommodate all available \glspl{RBG} and looping only through \gls{MUMIMO} layers to reduce computational complexity. These \gls{1L} and \gls{2L} designs are described in Appendix \ref{sec:rl_framework}. We describe our modified algorithms based on \gls{PPO} and \gls{SACD} in Appendices \ref{sec:on-policy-framework} and \ref{sec:off-policy-framework}, respectively. Performance comparisons are made against the baseline and the \gls{PF} greedy scheduler based on the exhaustive search defined in Appendix \ref{sec:reference_scheduler}.


All deep schedulers were trained and evaluated using a proprietary enterprise-grade \gls{3GPP}-compliant \gls{5GNR} system level simulator.
To assess deployment readiness, we further validated the trained models on commercial \gls{BTS} hardware under real-time conditions.
These tests confirmed that the proposed approach meets practical latency requirements, reinforcing its applicability to operational networks. Detailed results are omitted due to confidentiality.

\subsection{Evaluation}
The key simulation parameters are collected in Table \ref{tab:sys_params}, Appendix \ref{sec:detailed_params}, where the focus is on downlink non-contiguous (Type 1) allocation.
This enables unrestricted frequency-selective scheduling with \gls{RBG} granularity.
We assume the availability of sub-band \gls{CQI} and \gls{PMI} reports at the gNB and limit the maximum user-reported \gls{RI} to 2.
To assess the generalization capabilities of our deep schedulers, we evaluated performance via separate simulations for each of the two traffic models (\gls{FB} and FTP3 bursty), each employing 210 users to match the per traffic type scale used during training.

\subsection{Training}
Our on-policy and off-policy training methods employed a centralized approach, where a single \gls{RL} agent interacts with the multi-cell system-level simulator.
At each \gls{TTI}, the agent collects state-action-reward-next state tuples from all simulated \glspl{gNB} and updates a shared model.
This yields a generalized model that can be used independently by all \glspl{BTS}.
This design choice was motivated by its faster convergence and superior simulation performance compared to distributed training, while ensuring that the resulting actor models are directly deployable on real \gls{5G} \gls{BTS} hardware without additional fine-tuning.

To expedite training, models are trained on a reduced system bandwidth with fewer \gls{MUMIMO} layers, as detailed in Table \ref{tab:sys_params}, Appendix \ref{sec:detailed_params}.
This simplification was necessary to be able to train \gls{2L} models for 18 \glspl{RBG}.
Then, to promote generalization, we trained the models under mixed traffic conditions, utilizing a total of 420 users, where 210 users had \gls{FB} traffic, and 210 users had  FTP Model 3 traffic (see Table \ref{tab:sys_params}, Appendix \ref{sec:detailed_params}).
Carefully tuned hyperparameters for both methods are shown in Tables \ref{tab:ml_params} and \ref{tab:arch_params} of Appendix \ref{sec:detailed_params}.
Note that the replay buffer size for \gls{PPO} refers to the number of state-action pairs, in contrast to the traditional experience replay buffers.

For on-policy training, we adopt \gls{PPO} enhanced with expert guidance.
Specifically, we incorporate a \gls{JSD} loss term to align the agent’s policy with a \gls{PF} expert during early training, accelerating convergence and improving stability. 
This hybrid approach combines the robustness of \gls{PPO} with imitation learning benefits, enabling the agent to learn effective scheduling strategies under limited sample diversity.

On the off-policy \gls{RL} front, and as noted by \cite{pmlr-v139-ceron21a}, distributional \gls{RL} does not universally outperform non-distributional variants.
In our experiments, this effect was architecture-dependent: the \gls{1L} design, with its large multi-branch output layer and higher variance in action-value estimates, benefited from distributional critics, whereas the \gls{2L} design (making finer-grained decisions per \gls{RBG}) showed no measurable gain from distributional modeling.
Consequently, \gls{1L} was trained with \gls{DSACD}, while \gls{2L} used \gls{SACD}.

\begin{figure}[!ht]
\setlength\abovecaptionskip{0.0\baselineskip}
\centering
\includegraphics[width=5.0in]{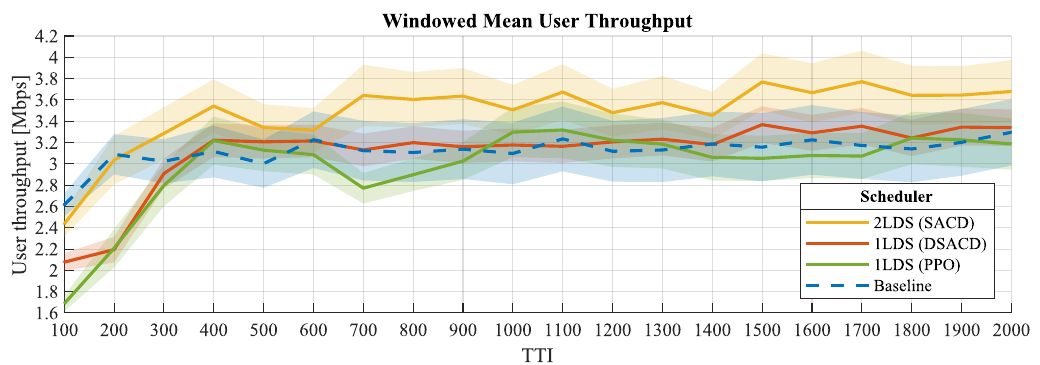}
\caption{Convergence of windowed mean user throughputs during the training of best performing \gls{1L} and \gls{2L} options with 95\% confidence interval. Due to higher entropy required by training \gls{2L} than \gls{1L} lower throughputs can be observed during the training.}
\label{fig_learning}
\end{figure}

Fig. \ref{fig_learning} illustrates smoothed learning curves for the best-performing models.
It should be mentioned that the target entropy values $\beta$ differed significantly between the \gls{1L} and \gls{2L} models.
Although \gls{2L} models performed best with a low $\beta$ of $0.4$, emphasizing the exploitation of learned strategies, \gls{1L} models required a high $\beta$ of $0.999$, potentially due to their limited capacity to benefit from increased exploration.
Additionally, \gls{1L} models, with their larger multi-branched output layers, needed more training samples to match the baseline scheduler's performance.
This convergence occurred around the 400th \gls{TTI}, at which point roughly 450,000 training $sars'$ tuples had been collected and used for training.
Training samples (state-action-reward-state tuples) are collected starting only after the 100th \gls{TTI} to allow the simulation to move past the initial transient phase while \gls{gNB} transmit buffers reach a representative state.
Specifically, considering the training parameters (Max UE layers $|L| = 4$ and $N_{RBG}=18$, see Table \ref{tab:sys_params}, Appendix \ref{sec:detailed_params}) and the 21-cell deployment, one sample is generated for each scheduling decision per layer, per \gls{RBG}, and per \gls{gNB}, resulting in $4\times18\times21=1512$ samples collected across the network in each subsequent \gls{TTI}.
This collection rate aligns with the accumulation of approximately $450,000$ tuples by the 400th \gls{TTI}.

The training performance of all models has been compared using geometric mean throughput.
This is the \gls{KPI} that best captures proportional fairness, aligning with our primary goal for \gls{FDS}/\gls{SDS} scheduling phases.
\gls{DSACD} was the only method to surpass the baseline scheduler in the \gls{1L} architecture, making it the preferred choice for this configuration.
For \gls{2L} architectures, \gls{SACD} outperformed all others.

\subsection{Assessment of Numerical Performance}
\begin{figure}[!ht]
\setlength\abovecaptionskip{0.0\baselineskip}
\centering
\begin{minipage}{2.6in}
\setlength\abovecaptionskip{0.0\baselineskip}
\includegraphics[width=2.6in]{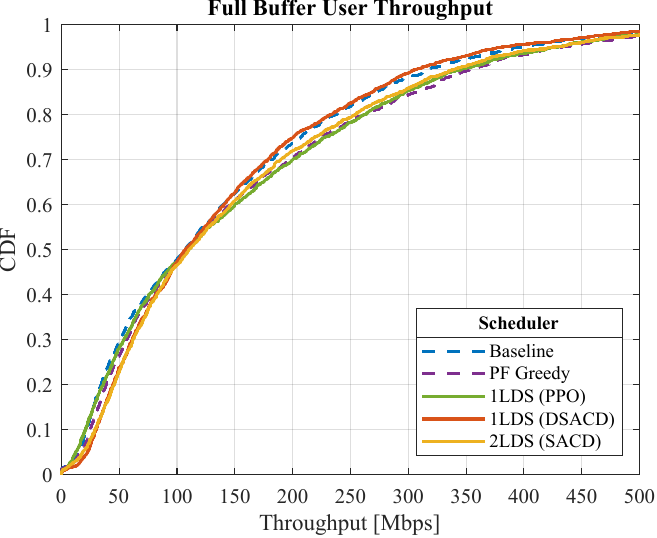}
\end{minipage}
\begin{minipage}{2.6in}
\setlength\abovecaptionskip{0.0\baselineskip}
\centering
\includegraphics[width=2.6in]{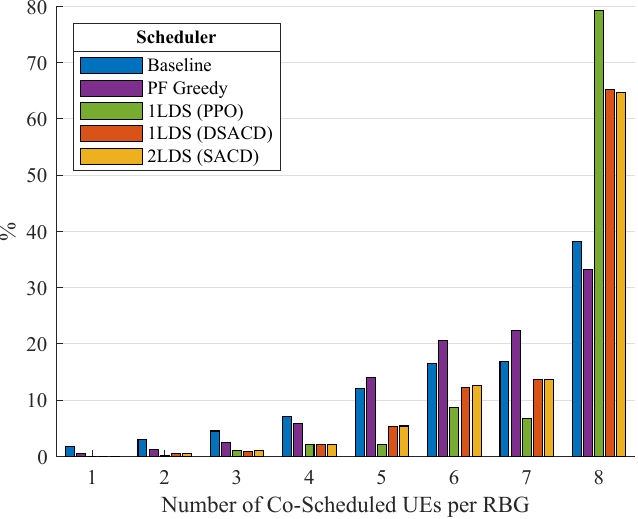}
\end{minipage}
\caption{User throughput distributions and co-scheduling efficiency with Full Buffer traffic.} 
\label{fig_fb_tput}
\end{figure}

\begin{figure}[!ht]
\centering
\begin{minipage}{2.6in}
\setlength\abovecaptionskip{0.0\baselineskip}
\centering
\includegraphics[width=2.6in]{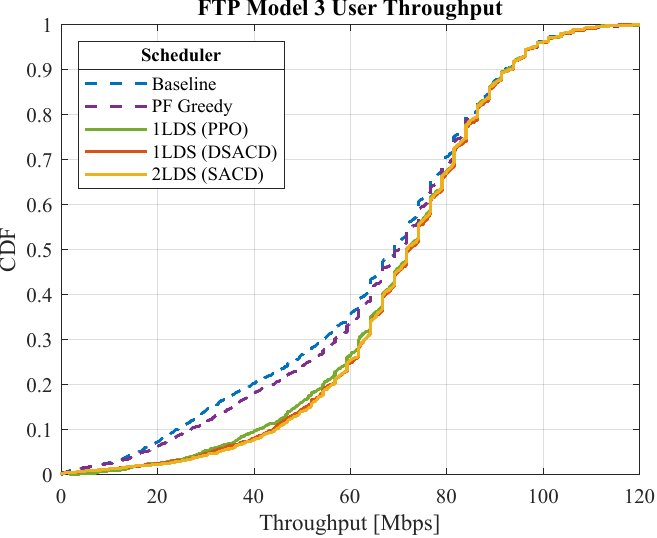}
\end{minipage}
\begin{minipage}{2.6in}
\setlength\abovecaptionskip{0.0\baselineskip}
\centering
\includegraphics[width=2.6in]{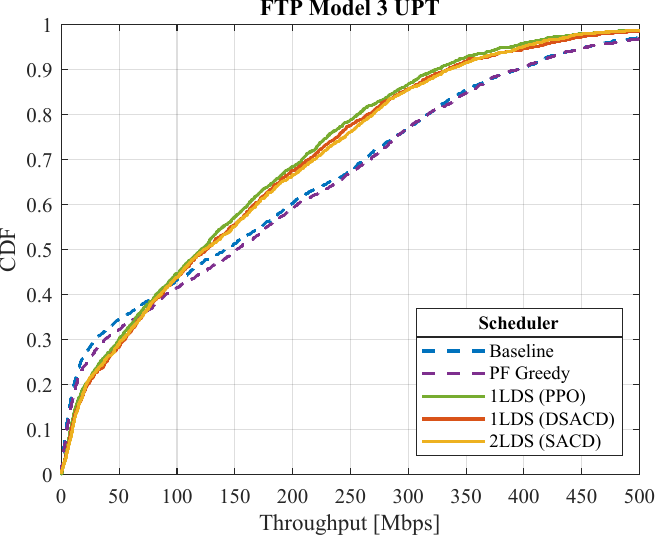}
\end{minipage}
\caption{User throughput and \gls{UPT} distributions with FTP Model 3 traffic. Most improvements are obtained among users in worst radio conditions.}
\label{fig_ftp3_tput}
\end{figure}

The final deep scheduler evaluations assessed the models under conditions significantly different from the training setup.
Specifically, the evaluations used the full 100 MHz bandwidth, twice the maximum number of \gls{MUMIMO} user layers ($|L|=8$ vs $|L|=4$), and homogeneous traffic (100\% FB or 100\% FTP3), contrasting sharply with the resource-constrained (6.6 MHz), lower order \gls{MUMIMO}, mixed traffic training environment (see Table \ref{tab:sys_params} in Appendix \ref{sec:detailed_params} for full details).
Therefore, the strong performance and gains over baseline schedulers observed in these evaluations (detailed below, see Figures \ref{fig_fb_tput}-\ref{fig_ftp3_tput}, Table \ref{tab:sim_result_summary}) serve as direct quantitative evidence substantiating the claimed generalization capabilities across these different operating conditions.
The results are collected from 10 different random simulation realizations.
To ensure comparable geometric mean values, zero-valued \gls{UE} throughputs are replaced with ones when calculating geometric means, as some random \gls{UE} drops result in a few \glspl{UE} being unable to receive data.

Figure \ref{fig_fb_tput} shows that all deep scheduler options perform well with full buffer traffic.
Notably, the \gls{2L} configuration provides better throughput overall.
In general, deep schedulers efficiently utilize spatial degrees of freedom by selecting \glspl{UE} for each \gls{RBG} and \gls{MUMIMO} user layer that can be co-scheduled with others while maintaining fairness.
Figure \ref{fig_fb_tput} also highlights a key advantage over heuristics: that deep schedulers achieve higher \gls{MUMIMO} co-scheduling efficiency.
This demonstrates the RL agents' ability to better navigate the complex combinatorial user pairing problem compared to the greedy, iterative nature of the baseline heuristics.
In contrast the \gls{1L} (PPO) approach may sometimes over-schedule a few \glspl{UE} for a single \gls{RBG}, leading to minor degradation in user throughputs compared to SAC-based options.

Given the discontinuous nature of FTP traffic, we use \gls{UPT} to evaluate network performance with bursty traffic models.
\gls{UPT}, defined as the throughput measured only when a \gls{UE} has downlink data to receive, is illustrated in Figure \ref{fig_ftp3_tput}.
The figure shows clear improvements in the fairness of the \gls{UPT} distribution achieved by deep schedulers.
A similar conclusion can be drawn from the classic user throughput CDF shown in Figure \ref{fig_ftp3_tput}.
The results reveal a particular strength under bursty traffic, where Table \ref{tab:sim_result_summary} and Figure \ref{fig_ftp3_tput} show dramatic improvements for cell-edge users (5th percentile) and fairness (geometric mean) compared to PF-based heuristics.
This suggests that the learned policies overcome limitations in heuristic approaches in balancing efficiency and fairness for disadvantaged users under dynamic load.
Table \ref{tab:sim_result_summary} summarizes the numerical throughput gains for separately simulated full buffer and bursty FTP model 3 traffic. All proposed deep schedulers achieve substantial gains, particularly for users in the worst radio conditions.
Notably, the least complex variant, \gls{1L} \gls{DSACD}, performs almost as well as the more complex \gls{2L} variant.
\begin{table}[h!]
\centering
\caption{Summary of worst 5th \%-ile, median, and geometric mean user throughput gains over the baseline with full buffer and FTP Model 3 traffic.}
\label{tab:sim_result_summary}
\begin{tabular}{cccc}
\toprule 
\textbf{Full Buffer} & \textbf{ 5th \%-ile } & \textbf{ Median } & \textbf{ Geomean } \\ 
\midrule 
\gls{PF} Greedy        & 11.1 \% & 1.3 \%  & 9.5 \% \\ 
\gls{1L} (PPO)   & -17.6 \% & 2.2 \%  & 6.4 \% \\
\gls{1L} (DSACD) & 46.7 \% & 0.2 \%  & 13.7 \% \\
\gls{2L} (SACD)  & 20.6 \% & 3.1 \%  & 15.6 \%\\
\midrule 
\textbf{FTP Model 3} & \textbf{ 5th \%-ile } & \textbf{ Median } & \textbf{ Geomean } \\ 
\midrule 
\gls{PF} Greedy        &  7.9 \% & 1.4 \%  & 3.5 \% \\ 
\gls{1L} (PPO)   &  85.5 \% &  3.5 \%  & 18.5 \% \\
\gls{1L} (DSACD) &  97.5 \% &  4.5 \%  & 18.3 \% \\
\gls{2L} (SACD)  &  106.3 \% &  3.8 \%  & 18.6 \% \\
\bottomrule 
\end{tabular}
\end{table}

Our computational complexity measurements (Appendix \ref{sec:complexity}, Table \ref{tab:execution_times}) indicate that \gls{1L} meets strict per-\gls{TTI} latency budgets, whereas \gls{2L} can struggle due to the $N_{RBG} \times L$ forward-pass requirement.

\section{Conclusion}
We presented practical \gls{RL}-based deep schedulers for 5G NR, introducing architectural and algorithmic innovations that enable frequency- and spatial-domain scheduling under strict latency requirements.
We did not rely solely on \gls{SoTA} \gls{RL}; instead, we introduced specific modifications to enhance its suitability for deep scheduler training.
For \gls{PPO}, incorporating expert guidance via \gls{JSD} loss represents a key enhancement over standard implementations.
Furthermore, we derived \gls{DSACD}, a novel \gls{SAC} variant, combining advances from state-of-the-art algorithms by integrating distributional critics for richer value representation and adaptive, state-aware entropy learning critical for stability with dynamic action spaces, along with tailored prioritized replay.
By employing a multi-branch output layer to schedule each \gls{MUMIMO} layer separately, DSACD provided the best trade-off between \gls{PF} throughput and computational complexity for frequency-selective massive MIMO scheduling.

Our analysis also revealed that while deep schedulers based on \gls{2L} architectures may struggle to meet \gls{5G} \gls{TTI} time limits, \gls{1L}-based designs offer a good balance between throughput performance and computational efficiency. This makes the \gls{1L} architectures a viable option for real-time applications, ensuring scalability to massive MIMO without compromising performance.

These findings underscore the potential of modified \gls{RL} algorithms in efficiently scheduling massive radio resources, paving the way for more sophisticated and resource-laden next-generation cellular networks.

\begin{ack}
The authors would like to thank their colleagues Ian Garcia, Hua Xu, Igor Filipovich, Simon Teferi, Morgan Cormier, Ngoc Duy Nguyen, Simon Goedecke, Andre Gille, Marian Beniac, Simo Aaltonen, Petri Ervasti, Silvio Mandelli, and Pavan K. Srinath for their feedback during numerous exchanges.
The European Union funds the work of B. Liu and A. Valcarce through the CENTRIC project (G.A no. 101096379).
\end{ack}

{
\small
\bibliographystyle{plainnat}
\bibliography{refs.bib}
}


\newpage
\appendix


\section{Previous Deep Scheduler efforts}
\label{sec:ds_history}

\begin{table}[H]
\centering
\caption{Deep Scheduler recent history}
\label{DS_history}

\begingroup
\setlength{\tabcolsep}{4pt}          
\renewcommand{\arraystretch}{0.95}   
\footnotesize                        

\begin{adjustbox}{max width=\textwidth}
\begin{tabular}{@{}L{1.3cm}C{0.5cm}C{0.5cm}L{1.5cm}L{2cm}L{1.6cm}L{1.6cm}L{1.3cm}L{1.4cm}L{1.6cm}@{}}
\toprule

\multirow{2}{*}{\textbf{Ref.}} & \multirow{2}{*}{\textbf{UL}} & \multirow{2}{*}{\textbf{DL}} & \multirow{2}{*}{\textbf{ML focus}} & \multirow{2}{*}{\textbf{Opt. target}} & \multicolumn{5}{c}{\textbf{Actor model}} \\
\cmidrule(lr){6-10}
 &  &  &  &  & \textbf{Algorithm} & \textbf{Input} & \textbf{Hidden} & \textbf{Output} & \textbf{Act. fun.} \\
\midrule

2020 \cite{9110842} &  & \checkmark & \gls{FDS} & bitrate fairness &
\acrshort{DDQN} & $2\cdot|U|$ & $2\times128$ & $|U|$ & ReLU \\

2021 \cite{9364885} &  & \checkmark & \gls{QoS} slicing & reliability &
\acrshort{AC}, \acrshort{PG} & $2\cdot|U|+1$ & N.K. & $N_{\mathrm{res}}$ & N.K. \\

2023 \cite{10247079} & \checkmark &  & \gls{FDS}, \gls{SDS} & bitrate fairness &
\acrshort{SAC}, KNN & $3\cdot|U|$ & $2\times64$ & $1$ & ReLU \\

2024 \cite{QoSdrama} &  & \checkmark & \gls{FDS} & packet loss rate &
\acrshort{PPO} & $3\cdot 6=18$ & $5\times64$ & $5$ & N.K. \\

2024 \cite{10329922} & \checkmark &  & MCS, $N_{\text{RBs}}$ & bitrate &
\acrshort{DDPG} & $2\cdot|U|$ & \makecell{$128,$ \\ $3\times256,$ \\ $128$} & $2$ & ReLU, $\sigma(x)$ \\

2024 \cite{Ericsson2024} &  & \checkmark & \gls{FDS}, \gls{SDS} & bitrate fairness &
\gls{DQN}, \gls{AB} & \makecell{$|U|\cdot N_{\mathrm{RBG}}$ \\ $\cdot D \cdot N_{tx}$} &
\makecell{$\sim$3.6M \\ params} & $N_{\mathrm{RBG}}\cdot N_a$ & ReLU \\

\bottomrule
\end{tabular}
\end{adjustbox}
\endgroup
\end{table}

\section{Framework for RL-based Radio Resource Schedulers}
\label{sec:rl_framework}

\begin{figure}[!ht]
\setlength\abovecaptionskip{0.0\baselineskip}
\centering
\includegraphics[width=5in]{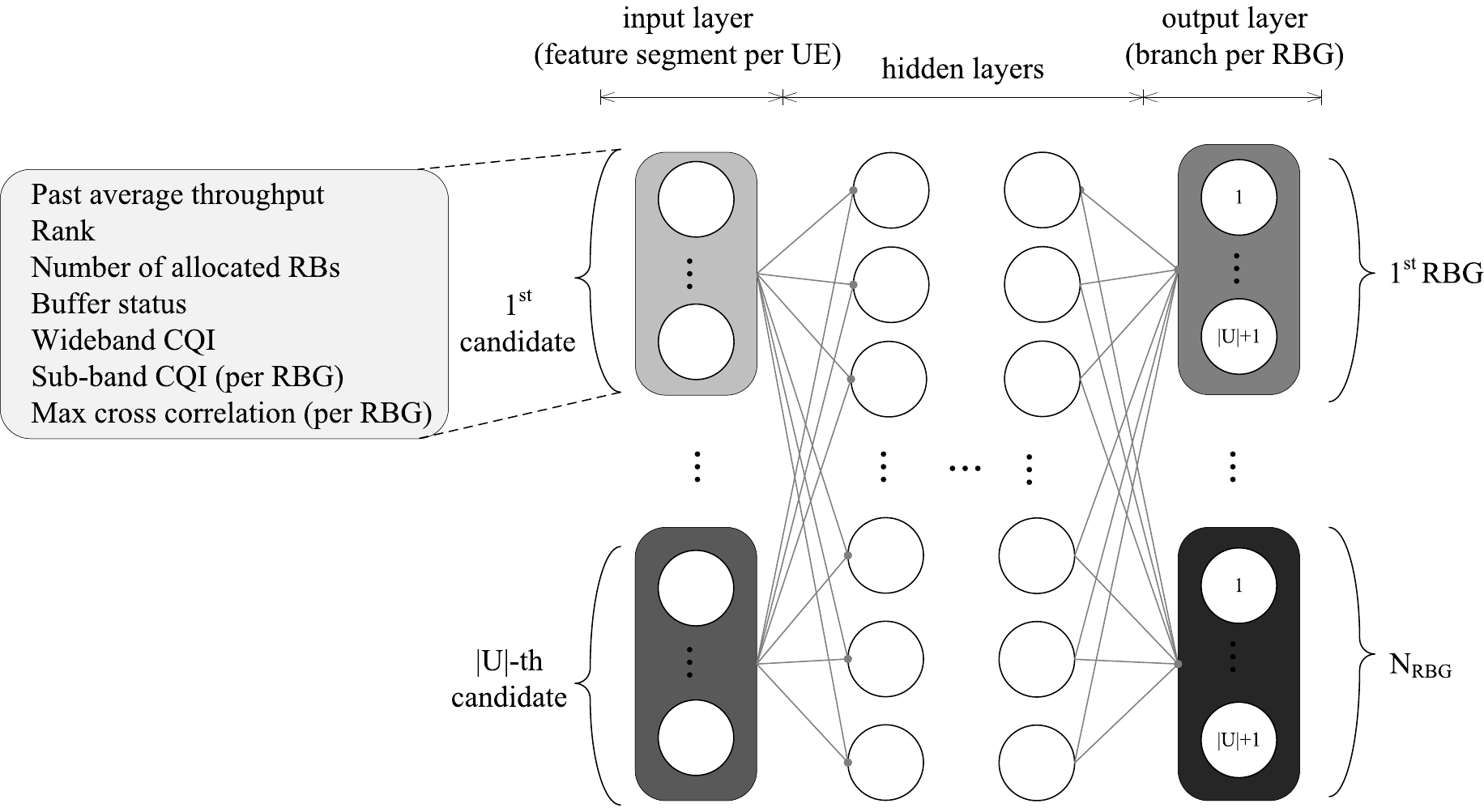}
\caption{\gls{1L} baseline neural network architecture, with feature segment blanking for variable \gls{UE} support. $|U|$ indicates maximum schedulable UEs per \gls{MUMIMO} user layer.}
\label{fig_nn_topo}
\end{figure}

To develop a practical deep radio resource scheduler that meets real-time \gls{BTS} constraints while maintaining or exceeding legacy scheduler \glspl{KPI}, we devised two \gls{ML}-based scheduler architectures.
The first, a \gls{1L}, optimizes execution time by providing scheduling decisions for all \glspl{RBG} per \gls{MUMIMO} user layer in a single forward pass, as illustrated in Fig. \ref{fig_nn_topo}.
In contrast, the second \gls{2L} approach decides the user allocation of one single \gls{RBG} per inference pass, requiring loops over both \glspl{RBG} and \gls{MUMIMO} user layers.
In both schemes, looping over the user layers is necessary.

Despite using larger input and final neural layers, the \gls{1L} design achieves a significant speedup due to fewer forward passes while managing to still keep \gls{ML} model dimensions rather small.
Concretely, \gls{1L} reduces the number of forward passes from $N_{\text{RBG}} \times |L|$ down to $|L|$, where $|L|$ denotes the maximum number of spatially co-scheduled users per \gls{RBG}, and $N_{\text{RBG}}$ is the total number of downlink \glspl{RBG} to be allocated.
This reduction is key in cellular systems with large bandwidths.
However, performing a single forward pass for all \glspl{RBG} and \gls{MUMIMO} layers would require a model that is too large for practical massive \gls{MIMO} implementation.

\subsection{Action Space}
For each decision stage, the deep scheduler selects a user from the set of candidate \glspl{UE} provided by the \gls{TD} scheduler or chooses not to allocate any user.
The action space is thus defined as $\mathcal{A} = \{1, \dots, |U|+1\}$, where $|U|$ is the number of candidate \glspl{UE}, and the additional action $|U|+1$ represents no allocation.
The implementation of a decision stage, however, differs between the \gls{1L} and \gls{2L} architectures. 
In the \gls{1L} approach, a single decision stage corresponds to a complete user-layer.
The scheduler performs one forward pass of the neural network to determine the action (user allocation or no allocation) for all \glspl{RBG} within that specific user-layer simultaneously.
Conversely, in the \gls{2L} architecture, a decision stage is implicitly equivalent to the allocation of an \gls{RBG} within a user-layer.
The scheduler iterates through each \gls{RBG} and each user-layer, making an independent allocation decision at each step.
Action masking is employed in both architectures to prevent invalid allocations, such as assigning already scheduled \glspl{UE} to the same \gls{RBG} or considering empty candidate inputs.
In this framework, the deep scheduler makes decisions at each \gls{TTI}, assigning a user (or no user) to each \gls{RBG} per \gls{MUMIMO} layer. This process implicitly defines a time-frequency resource allocation pattern over consecutive \glspl{TTI}.

\subsection{State Space}
\label{sec:state_space}

The deep scheduler inputs are grouped into so-called state vectors, whose features have been chosen for their sufficiency in scheduling decision-making and low computational demands. 
Moreover, only values that the \gls{5G} standard supports, are already available in \gls{BTS}, and do not require additional signaling overhead are used as input features.
Since features are collected per \gls{UE}, they effectively construct what we call a \emph{\gls{UE} feature segment}, and the state vector is therefore built out of $|U|$ \gls{UE} feature segments. Each segment includes normalized values of \gls{UE}'s past average throughput, rank, number of allocated \glspl{RB}, wideband \gls{CQI}, sub-band \gls{CQI} (per \gls{RBG}) and max cross correlation (per \gls{RBG}).

Next we describe the input features in detail, although for clarity, and omit the \gls{TTI} index as follows: A prime denotes the variable at the next state, e.g., $x'$, and a double prime denotes the variable at the previous state, e.g., $x''$.

\begin{itemize}
   \item \textbf{Normalized Past Averaged Throughput ($\hat{R}_u$):}
   The past averaged throughput for user $u$ is calculated using exponential smoothing:
   $$R_u = (1 - \epsilon) p_u + \epsilon R_u'',$$
   where $p_u$ is the instantaneous throughput, $R_u''$ is the previous averaged throughput, and $\epsilon$ is the forgetting factor. The normalized past averaged throughput is then:
   $$\hat{R}_u = \frac{R_u}{R_{\text{max}}},$$
   with $R_{\text{max}}$ being the maximum observed past averaged throughput.

   \item \textbf{Normalized Rank of UE ($\hat{h}_u$):}
   The rank of user $u$, denoted by $h_u$, represents the \gls{UE}'s recommendation for the number of transmission layers. We normalize it as:
   $$\hat{h}_u = \frac{h_u}{2},$$
   The $2$ in the denominator stems from the maximum \gls{UE} rank (see \Cref{tab:sys_params}).

   \item \textbf{Normalized Number of Already Allocated \glspl{RBG} ($\hat{d}_u$):}
   $d_u$ denotes the number of \glspl{RBG} allocated to user $u$ during the previous iteration of the scheduling loop across the \gls{MUMIMO} layers. This is then normalized as:
   $$\hat{d}_u = \frac{d_u}{d_{\text{max}}},$$
   where $d_{\text{max}}$ is the bandwidth-dependent total number of \glspl{RBG} across the full bandwidth.

   \item \textbf{Normalized Downlink Buffer Status ($\hat{b}_u$):}
   The downlink buffer status for user $u$, denoted by $b_u$, is normalized as:
   $$\hat{b}_u = \frac{b_u}{b_{\text{max}}},$$
   where $b_{\text{max}}$ is a predefined maximum possible buffer size.

   \item \textbf{Normalized Wideband \gls{CQI} ($\hat{o}_u$):}
   Min-max scaling is used to scale the wideband \gls{CQI} for user $u$ to the $[0, 1]$ range as follows:
   $$\hat{o}_u = \frac{o_u - o_{\text{min}}}{o_{\text{max}} - o_{\text{min}}},$$
   where $o_u$ is the current wideband CQI, and $o_{\text{min}}$ and $o_{\text{max}}$ are the minimum and maximum possible CQI values, respectively.
\end{itemize}

\textbf{Per-\gls{RBG} User-Specific Input Features:}

\begin{itemize}
   \item \textbf{Normalized Sub-band CQI ($\hat{g}_{m,u}$):}
   The sub-band CQI for user $u$ on \gls{RBG} $m$, denoted by $g_{m,u}$, is normalized as:
   $$\hat{g}_{m,u} = \frac{g_{m,u}}{\bar{g}_{m,u}},$$
   where $\bar{g}_{m,u}$ is a predefined normalization scalar.
   
   \item \textbf{Max Cross-Correlation in User Pairing ($\rho_{m,u}$):}
   The maximum user pair cross-correlation between already scheduled users and the new candidate $u$ on that \gls{RBG} $m$ is computed. This pre-calculated value is based on the precoder matrices $P_{m,u}$ and $P_{m,c}$ (for users $u$ and $c$ co-scheduled on \gls{RBG} $m$)  and is given by:
   $$\rho_{m,u} = \max\{\kappa_{m,u,1}, \kappa_{m,u,2}, \dots, \kappa_{m,u,N_{m,l}}\},$$
   where $\kappa_{m,u,c} = \max\{\sum_i |[P_{m,u}^H P_{m,c}]_{i,j}|,$ for $j \in \{1, 2, \dots, h_c\}\}$, $h_c$ is the rank of user $c$, $N_{m,l}$ is the number of co-scheduled users on \gls{RBG} $m$ at layer $l$, and $[\cdot]_{i,j}$ denotes the element at the $i$-th row and $j$-th column of a given matrix.
\end{itemize}

All inputs come from standard CSI and buffer reports. No extra uplink signaling is introduced.
Having defined the input features, we can now specify the state vectors for the different architectural variants.
For the fast \gls{1L} architecture (Fig. \ref{fig_nn_topo}), the state vector includes five features $(\hat{R}_u, \hat{h}_u, \hat{d}_u, \hat{b}_u$, $\hat{o}_u)$ and two RBG-specific features ($\hat{g}_{m,u}, \rho_{m,u}$) for each candidate \gls{UE}.
The size of this state vector is $|U| \times (5 + 2N_{\text{RBG}})$.
On the other hand, the \gls{2L} scheduler, designed for throughput performance, executes one forward pass per \gls{RBG} and receives \gls{RBG}-specific inputs.
These inputs include the seven previously defined features plus an eighth feature for the mean precoder cross-correlation, as well as the number of co-scheduled users so far.
The input vector size for the \gls{2L} scheduler is thus $|U| \times 8 + 1$.

The defined state features enable adaptation to key network dynamics.
Channel variations, including fading and multi-cell interference, are implicitly captured via the comprehensive \gls{CSI} inputs (\glspl{CQI}, rank, correlation).
The local traffic load is reflected in the buffer status and past throughput features.
Although the current design handles \gls{FD}/\gls{SD} scheduling across different traffic models, explicitly incorporating fine-grained QoS parameters remains a future work. This is because traditionally \gls{QoS} optimization is seen as a job of scheduling candidate selection, which is outside the scope of this study.
Considering all these neural architectures and input feature vectors, there exist multiple ways to train them. In the following sections, we describe different training algorithms and their performance.

\subsection{Optimization Goal}
All of the deep schedulers shown here share one common objective: Maximizing the long-term geometric mean of user throughput.
This is a commonly used \gls{KPI} of radio schedulers, which captures both network capacity and fairness of resource allocation. It is defined as follows:
\begin{align}
\label{eqn:OptGoal}
    G = \bigg{(}\prod_{u=1}^{|U|}p_u\Bigg{)}^{1/|U|}.
\end{align}

 \section{On-Policy Deep Scheduler Training}
\label{sec:on-policy-framework}
Although pre-trained \gls{DRL} schedulers may be sufficient in most cellular scenarios, niche applications like military deployments require rapid adaptation to unpredictable radio environments.
In these safety-critical situations, the stability and cautious exploration of on-policy \gls{RL} methods are advantageous.

\subsection{On-policy Framework with Expert Guidance}
\begin{figure}[!ht]
\centering
\includegraphics[width=0.98\columnwidth]{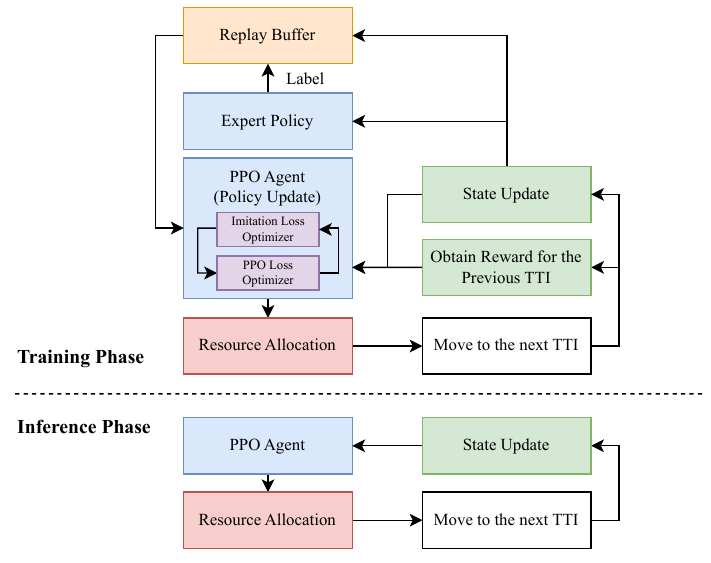}
\caption{Overall Framework of 1LDS - PPO}
\label{fig:DeepScheduler_PPO}
\end{figure}

On-policy \gls{RL} algorithms like \gls{PPO} are favored for their ease of implementation and stability. 
However, their reliance on the current policy for environment interaction can lead to slower convergence due to limited sample diversity.
Addressing the pertinent challenge of enhancing sample efficiency and accelerating learning in complex \gls{RL} tasks like scheduling, our proposed \gls{1L}-based framework integrates an \emph{expert policy} to guide the \gls{PPO} agent during training (similar to the guide policy described in \cite{Uchendu2022JumpStartRL}).
This expert policy, a \gls{PF} scheduler in our case, acts as a source of additional training data, and is illustrated in Fig. \ref{fig:DeepScheduler_PPO}.

The training phase involves four key stages:
\begin{itemize}
    \item \textbf{State and reward observation}: Following traditional reinforcement learning, the input state for the actor and critic networks is updated at each new \gls{TTI}. Additionally, the reward from the previous \gls{TTI} is collected and saved into an experience buffer for the \gls{PPO} agent to update its policy. 
    
    \item \textbf{Expert label generation}: Based on system requirements and computational resources, an expert policy generates labels to guide the \gls{PPO} agent.
    
    \item \textbf{Policy update}: Once the experience buffer reaches its capacity, the \gls{PPO} agent updates its policy using the collected data and expert labels. This is achieved through back-propagation and optimization of a pre-defined loss function.
    
    \item \textbf{Resource allocation}: The deep scheduler returns a vector of \gls{UE} candidate indices for each \gls{RBG}. After that, the candidates vector is sampled and the resource allocator assigns the \gls{RBG} to the chosen \gls{UE}. Modulation and coding schemes are selected after all \glspl{RBG} have been allocated.
\end{itemize}

This approach leverages the strengths of both on-policy \gls{RL} and expert knowledge, leading to faster convergence and improved scheduling performance in a practical setting.
The other fundamental components of training a \gls{1L}-\gls{PPO} are detailed next.
The complete training procedure is formally outlined in Algorithm \ref{alg:ppo_expert}, Appendix \ref{sec:algorithms}.

\graphicspath{{./figures/}}

\subsection{Fundamental Components}

\subsubsection{Action Space}\label{sec:ActionSpace}
The final layer of the 1LDS-PPO actor network outputs a vector $\mathbf{z} \in \mathbb{R}^{(|U|+1) N_\text{RBG}}$, which is then reshaped into $\hat{\mathbf{z}} \in \mathbb{R}^{N_\text{RBG} \times (|U|+1)}$.
Let $\text{softmax}(X, d)$ be a softmax function to be applied to the $d$-th dimension of an input matrix $X$.
The softmax function is applied along the second dimension (representing \gls{UE} candidates) to obtain a probability distribution $\Tilde{\mathbf{z}} \in \mathbb{R}^{N_\text{RBG} \times (|U|+1)}$ over the \glspl{UE} for each \gls{RBG}:
$\Tilde{\mathbf{z}} = \text{softmax}(\hat{\mathbf{z}}, 2)$,
where $\sum_{u=1}^{|U|+1}\Tilde{z}_{m,u} = 1$ for $m \in \{1,2,...,N_\text{RBG}\}$.

From this distribution, the action vector $\mathbf{a} = [a_1, a_2, ..., a_{N_\text{RBG}}]^\text{T}$ is sampled, with each element $a_m \in \{1,2,...,|U|+1\}$ representing the index of the chosen \gls{UE} for \gls{RBG} $m$.
If the selected index does not correspond to a valid user, the \gls{RBG} remains unassigned.

In summary, the action determination process involves:

\begin{itemize}
    \item Obtaining the output vector $\mathbf{z}$ from the actor network.
    
    \item Reshaping $\mathbf{z}$ into $\hat{\mathbf{z}}$.
    
    \item Applying the softmax function to obtain the probability distribution $\Tilde{\mathbf{z}}$.
    
    \item Sampling the action vector $\mathbf{a}$ from $\Tilde{\mathbf{z}}$. 
\end{itemize}

\subsubsection{Reward Function}
The reward function is designed to balance two key objectives: maximizing the geometric mean of user throughput (for fairness) and maximizing the performance gain from \gls{MUMIMO}.
This approach encourages the scheduler to learn a policy that efficiently allocates resources while exploiting the benefits of \gls{MUMIMO}.
In a \gls{1L} architecture, the deep scheduler loops over the \gls{MUMIMO} user layers, and the reward for the $l$-th user layer and $m$-th \gls{RBG}, denoted by $r_{m,l}$, is defined as
\begin{align}\label{eq:ppo_rwd}
    r_{m,l} = \begin{cases}
        P \cdot v_m & \text{for} \ l = 1 \\
        k \cdot v_m & \text{otherwise}
    \end{cases},
\end{align}
where
\begin{align}
\label{eq:GM_throughput}
    P =  \frac{1}{G_\text{max}}\cdot G
\end{align}
refers to the geometric mean of user throughput at the current TTI, $G_{\text{max}}$ is a training-only normalizer, without dependencies at inference. We mitigate scale-shift by training under mixed traffic and test under stronger settings ($100$ MHz, $L=8$).
In addition,
\begin{align}
v_m = \begin{cases}
      -1 \quad \ & \text{if a better allocation choice exists} \  \\
      1 \qquad \quad \; \; & \text{otherwise}
   \end{cases}
\end{align}
denotes the result of a greedy search to determine whether another resource allocation exists, whose \gls{PF} metric exceeds the one selected by the deep scheduler.
$k$ is a factor that adjusts the effect of greedy search during user pairing.

We consider the greedy search indicator $v_m$ to determine whether an alternative \gls{UE} allocation could enhance the total \gls{PF} metrics. This indicator captures the trade-off between overall throughput and individual user fairness.
By incorporating throughput fairness and \gls{PF} metrics during user pairing into the reward function, the scheduler is guided to learn a policy that efficiently allocates resources while leveraging the spatial multiplexing capabilities of \gls{MUMIMO}.
Following conventional reinforcement learning training architecture, the reward is computed after each iteration of the user layer.

Furthermore, consistent with standard RL timing (see Fig. \ref{fig:DeepScheduler_PPO}), the reward $r_{m,l}$ corresponding to actions taken during TTI $t$ is computed based on the instantaneous throughput $p_u$ realized in that interval (which determines $G$ used in $P$), but this reward value becomes available for agent updates at the start of TTI $t+1$.
This inherent delay ensures the required throughput outcome from TTI $t$ is known before the corresponding reward is computed, resolving any apparent conflict with the timing of final MCS selection which occurs after allocations for TTI $t$ are complete.

\subsubsection{Training Data Augmentation}\label{sec:TrDataAugment}
To enhance generalization and accelerate training, we shuffle the \gls{UE} feature segments within the input state (see Fig. \ref{fig_nn_topo}).
This generates additional experience samples for the scheduler model to learn from while preserving the reward and swapping the \gls{UE} candidate indices according to the permutation order. 
The number of permutations, $N_{\Pi}$, is a pre-set hyperparameter.
Each new experience tuple saved in the buffer undergoes $N_{\Pi}$ permutations, effectively multiplying the available training samples.

An experience replay buffer $\mathcal{D}_\text{expert}$ with ample memory is also employed to store input state-action pairs from the expert policy.
This buffer provides the learning agent with diverse expert demonstrations to learn from, further improving its performance.
Unlike traditional off-policy replay buffers, which store rewards, next states, and termination flags, this buffer focuses solely on current states and corresponding expert actions.

\subsubsection{Loss Function with Expert Guidance}
To accelerate convergence and benefit from expert knowledge, we introduce an additional loss term to the standard \gls{PPO} loss function, encouraging the on-policy \gls{DRL} agent to mimic the expert policy.
The expert policy generates the labels for the scheduling decisions that the \gls{PPO} agent can learn from.
This term is based on the Jensen-Shannon Divergence (JSD), which measures the similarity between probability distributions.
JSD is symmetric, robust to zero probabilities (possible in radio scheduling), and bounded between $0$ and $1$, offering a normalized dissimilarity metric.

Specifically, we calculate the JSD loss, $\mathcal{L}_\text{JSD}$, between the actor network's output and the expert policy's label:
\begin{align}
    \mathcal{L}_\text{JSD} &= \frac{1}{2}\sum_{a_u}^{\{1,2,...,|U|+1\}}\pi_{\theta_{\text{PPO}}}(a_u|s)\text{log}\frac{\pi_{\theta_{\text{PPO}}}(a_u|s)}{\pi_\text{expert}(a_u|s)} \label{eq:JSD_loss} \\ \notag & + \frac{1}{2}\sum_{a_u}^{\{1,2,...,|U|+1\}}\pi_\text{expert}(a_u|s)\text{log}\frac{\pi_\text{expert}(a_u| s)}{\pi_{\theta_{\text{PPO}}}(a_u|s)} 
\end{align}
where:
\begin{itemize}
    \item $\pi_{\theta_{\text{PPO}}}(a_u|s)$ is the probability of the PPO agent taking action $a_u$ given state $s$.
    \item $\pi_{\theta_{\text{expert}}}(a_u|s)$ is the probability of an expert policy taking action $a_u$ given state $s$.    
\end{itemize}
To help the agent explore and converge to a policy better than the expert, the traditional PPO loss is employed and defined as:
\begin{align}
    \mathcal{L}_\text{PPO} &= L_\text{value} - L_\text{policy} - \xi L_\text{entropy}, \label{eq:ppo} \\
    L_\text{value} &= \frac{1}{2}\mathbb{E}\big{\{}(V(s) - \hat{J})^2\big{\}}, \label{eq:l_value}\\
    L_\text{policy} &= \mathbb{E}\bigg{\{}\text{min}\big{(}\varrho(\theta)\hat{A}, \text{clip}(\varrho(\theta), 1-\epsilon, 1+\epsilon)\hat{A}\big{)}\bigg{\}}, \label{eq:l_policy}\\ 
    L_\text{entropy} &= H(\mathbf{z}|s;\pi_{\theta_{\text{PPO}}}) \label{eq:l_entropy}
\end{align}
where:
\begin{itemize}
    \item $\xi$ is a coefficient controlling the strength of entropy regularization, promoting exploration.
    \item $H(\cdot)$ is the statistical entropy of the current policy.
    \item $\hat{A}$ is the advantage value, estimated using a \gls{GAE} \cite{gae}.
    \item $\varrho(\cdot) = \frac{\pi_{\theta_\text{PPO}}}{\pi_{\theta_\text{PPO}'}}$ is the ratio of probabilities between the updated policy ($\pi_{\theta_\text{PPO}}$) and the old policy ($\pi_{\theta_\text{PPO}'}$). The overall policy is calculated as the product of individual action probabilities across all \glspl{RBG}: $\pi_{\theta_\text{PPO}} = \prod_{i=1}^{N_\text{RBG}}\pi_{\theta_\text{PPO},i}$.
    \item $\epsilon$ is the clipping parameter to constrain policy updates and maintain stability.
    \item $V(\cdot)$ is the estimate of the state value function.
    \item $\hat{J}$ is the target return, computed as the cumulative sum of discounted rewards.
\end{itemize}
We train the agent using alternating optimization.
In each training step, the standard PPO loss ($\mathcal{L}_\text{PPO}$) is first minimized using batches of standard agent experiences.
Then, the $\mathcal{L}_\text{JSD}$ loss is minimized using expert state-action pairs sampled from a dedicated replay buffer, thereby incorporating expert guidance.
This alternating process allows the agent to learn simultaneously from its own experience (via PPO) and from the expert's guidance (via JSD), aiming for faster convergence and potentially better performance than using PPO alone.
Note also that two buffers are used: a small, temporary one for collecting the \gls{PPO} update batch (size $M=128$ as per Table \ref{tab:ml_params}) and a larger, persistent one for storing expert demonstrations for the \gls{JSD} loss (size $4000$ as per Table \ref{tab:ml_params}, Appendix \ref{sec:detailed_params}).
 \section{Off-Policy Deep Scheduler Training}
\label{sec:off-policy-framework}
Off-policy \gls{RL} methods offer a distinct advantage over their on-policy counterparts by allowing agents to learn from experiences generated by a different policy.
This enables efficient utilization of past scheduling decisions stored in a replay buffer, facilitating faster and more robust learning.
In radio resource scheduling, off-policy methods are particularly appealing when exploration and stability are critical.

This section investigates and combines several off-policy \gls{RL} advancements suitable for our radio resource allocation problem.
We first consider \gls{SAC} \cite{pmlr-v80-haarnoja18b}, which is proven to be a rather effective and sample efficient algorithm due to its entropy regularization.
However, as the original \gls{SAC} algorithm is designed for continuous action spaces, we adopt a modified version, \gls{SACD} \cite{christodoulou-sacd}, instead.
This variant is better suited for our discrete action space, with its critic networks estimating Q-values for all actions simultaneously, allowing the actor to learn from the entire Q-value distribution for a given state.
Furthermore, since nearby scheduling decisions are not necessarily correlated (especially with shuffled state feature segments and a correspondingly branched output layer, see \Cref{fig_nn_topo}), the Gaussian policy of the original \gls{SAC} is less suitable than the individual action outputs of \gls{SACD}, similar to traditional \gls{DQN}-based algorithms.

Finally, we explore distributional \gls{RL}, by extending \gls{SACD}, with distributional critic networks, resulting in \gls{DSACD}. With both \gls{SAC} variants, we learn a policy ${\pi}_{\theta}$ and two Q-functions $Q_{\phi 1}$ and $Q_{\phi 2}$, along with their soft-updated targets $\bar{Q}_{\bar{\phi} 1}$ and $\bar{Q}_{\bar{\phi} 2}$.

\subsection{Distributional Soft Actor-Critic Discrete}
A key problem in distributional \gls{RL} \cite{bdr2023} is that it increases the ML model size and computational complexity. This is not desired for models intended for real-time systems such as \glspl{BTS}. To capture the benefits of distributional learning and \gls{SoTA} maximum entropy \gls{RL} for discrete action spaces, we modify the online and target critic networks to be distributional while keeping the actor (policy) network unchanged. This allows us to train a model that combines maximum entropy \gls{RL} with distributional \gls{RL} and yield an actor model (policy network) that is no more complex to execute than the original \gls{DQN} \cite{Mnih13}.

To transform the critic networks into distributional ones, we apply two key modifications proposed in \cite{dabney_qr-dqn} to the \gls{SACD} critic networks. First, we expand the critics' output layers by a factor of $N$, a hyperparameter that determines the number of quantiles used to approximate the distribution of Q-values. This results in an output layer size of $|A| \times N$, where $|A|$ is the size of the action space.
This expansion allows the network to represent a distribution of Q-values for each action, providing a richer representation of the value function.
Second, we replace the \gls{MSE} loss used in the original \gls{SACD} critics with the quantile Huber loss.
This loss is more robust to outliers than \gls{MSE}, and is given by:
\begin{align} \label{eq:huber}
\begin{split}
    L(x) = \begin{cases}
      \dfrac{1}{2}x^2 & \text{for } |x|<k \\
      k(|x|-\dfrac{1}{2}k) & \text{otherwise}
    \end{cases}
\end{split},
\end{align}
where $k$ is a hyperparameter we set to $k=1$.

The quantile Huber loss, an asymmetric variant of the Huber loss is given by:
\begin{align} \label{eq:quantile_huber}
\begin{split}
   \rho_{\tau}^{k}(x) = |\tau - \delta_{\{x<0\}}| L(x)
\end{split},
\end{align}
where $\delta$ denotes a Dirac delta function and $\tau$ represents the quantile level, with $N$ quantiles evenly spaced between 0 and 1 as $\tau_{n} = \dfrac{n}{N}, \text{for } n=1,...,N$.
This loss function creates $N$ quantiles per action in the output layers of the critic networks, allowing them to represent a distribution of Q-values instead of a single point estimate.
This distributional representation captures uncertainty in the value estimation and can lead to improved performance in \gls{RL} tasks.

For our \gls{DSACD} algorithm, the error $x_t$ for each quantile $n=1,...N$ is defined as:
\begin{align} \label{eq:critic_loss}
\begin{split}
   x_{\phi,n}=q_{\phi,n}(s,a) - y_{\bar{\phi}}(r(s,a),{s'})
\end{split},
\end{align}
where $q_{\phi,n}(s,a)$ is the $n$-th quantile output of the critic network for state $s$ and action $a$, and the target value is calculated as:
\begin{align} \label{eq:critic_target}
\begin{split}
   y_{\bar{\phi}}(r(s,a),s') = 
   r(s,a) + \\
   \gamma \pi_{\theta}({s'},{a'}) \bigl(\bar{Q}_{\bar{\phi}}({s'},{a'}) - 
   \alpha \text{ log}(\pi_{\theta}({s'},{a'})) \bigl), \\ 
   {a'}\sim\pi_{\theta}(\cdot|{s'}).
\end{split}
\end{align}

Here, $\gamma$ is the discount factor, $r(s, a)$ is the reward, $\alpha$ is the learned entropy regularization coefficient, and $\pi_{\theta}({s'},{a'})$ is the probability of taking action $a'$ in state $s'$ according to the actor network.
The term $\bar{Q}_{\bar{\phi}}(s', a')$ represents the expected value from the target critic network for state $s'$ and action $a'$.
Note that action $a'$ for the next state $s'$ is drawn from the probability distribution given by the actor-network.
The target value $\bar{Q}_{\bar{\phi}}({s'},{a'})$ is obtained by averaging over all $N$ quantiles in the next state $s'$ for the drawn action $a'$, as follows: 
\begin{align} \label{eq:Q_target}
\begin{split}
   {Q}_{\phi}({s},{a}):=\sum_{n=1}^{N} \dfrac{1}{N} {q}_{\phi,n}({s},{a})
\end{split}.
\end{align}
This distributional approach captures the uncertainty in future rewards and allows for a more robust and nuanced learning process.

The actor-network is trained following the principles of \gls{SACD} \cite{christodoulou-sacd}.
During training, a softmax function is applied to the output layer (separately for each output branch for \gls{1L} approach) to convert the discrete outputs into a probability distribution over actions. However, to simplify inference and reduce complexity, the softmax operation can be omitted after training, allowing greedy action selection using argmax.
The main difference from the original \gls{SACD} \cite{christodoulou-sacd} lies in handling the distributional Q-values from the critic networks. Since the critics output a distribution of Q-values for each action, we compute the mean Q-value across all $N$ quantiles for each action.
This transforms the quantile outputs into a format comparable to the actor network's probability distribution.
To mitigate Q-value overestimation, \gls{SAC} utilizes two critic networks.
The policy objective function, which guides the actor's learning, is defined as:
\begin{align} \label{eq:policy_objective}
\begin{split}
   J_{\pi}(\theta) = \mathbb{E}_{s\sim D}\big{[}\pi_{\theta}(s)\odot[\alpha \text{ log}(\pi_{\theta}(s)) - \min_{j=1,2}Q_{\phi j}(s)]\big{]}
\end{split},
\end{align}
where $D$ is the replay memory and the array of mean Q-values $Q_{\phi j}(s)$ from critic network $j$ are obtained as described previously. 
It should be noted that the min operation selects the smaller Q-value from the two critics for each action, promoting conservative estimates.
This objective function guides the gradient descent process across all actions in the discrete action space. While the Q-networks are updated based on the actions taken, the policy is trained using all action probabilities and their corresponding Q-values, ensuring comprehensive learning from the state-action value distribution.

\subsection{Learning SACD Entropy with Action Masking}
To ensure the selection of valid scheduling candidates, we employ action masking to exclude candidates already scheduled for specific \glspl{RBG}.
Additionally, the number of valid actions may vary across TTIs due to the dynamic nature of \gls{TD} scheduling.

In \gls{SACD}, the entropy regularization parameter $\alpha$ is learned, and the entropy target depends on the size of the action space. Instead of using a static target value $\bar{H}$ as in \cite{christodoulou-sacd}, we adopt a state-specific target in the entropy objective:
\begin{align} \label{eq:enropy_objective}
\begin{split}
   J(\alpha) = \mathbb{E}_{s\sim D}[\pi_{\theta}(s)[\alpha(\text{log}(\pi_{\theta}(s)) + \bar{H}(s))]]
\end{split},
\end{align}
where the state-specific target is derived from the number of valid actions $|A(s)|$ in the current state:
\begin{align} \label{eq:entropy_H}
\begin{split}
    \bar{H}(s) &= -\beta \cdot \text{log} \left( \frac{1}{|A(s)|} \right).
\end{split}
\end{align}
Here, $\beta$ is a hyperparameter controlling the target entropy.
While the default value of $\beta=0.98$ from \cite{christodoulou-sacd} performs well, we further optimized our deep scheduler by tuning $\alpha$ based on the number of valid actions and searching for a more optimal $\beta$ (see values in Table \ref{tab:arch_params}, Appendix \ref{sec:detailed_params}).

Masks are stored in the replay memory alongside states, enabling us to determine the number of valid actions during training.
Since the actor produces probabilities for all actions, including invalid ones, the probability distribution is adjusted by setting invalid action probabilities to zero and renormalizing the remaining probabilities to sum to one.
Finally, the entropy is updated using gradient descent, averaging over all actions:
\begin{align} \label{eq:entropy_gradient}
\begin{split}
   \alpha\leftarrow\alpha-\dfrac{1}{|A(s)|}\sum_{a}^{A} \pi_{\theta}(s,a)[\text{log}(\pi_{\theta}(s,a))-\bar{H}(s)]
\end{split},
\end{align}
where the logarithm is applied element-wise to the probability distribution of the discrete action space.

\subsection{Prioritized Experience Replay}
Replay memory prioritization, based on the method proposed in \cite{replay_prio}, originally generated bias toward transitions with higher temporal difference (TD) errors in \glspl{DQN}.
However, in our \gls{DSACD} algorithm, we have two Q-networks.
Therefore, we adapt the prioritization strategy \cite{replay_prio}.
The priority $\delta_i$ assigned to a transition $i$ is based on the average absolute TD error across the critic networks and quantiles, ensuring transitions with larger estimation errors are prioritized:
\begin{equation} \label{eq:per_priority}
\delta_i = \frac{1}{2N}\sum_{j=1}^{2}\sum_{n=1}^{N}|x_{\phi j,n,i}| + \epsilon,
\end{equation}
where $x_{\phi j,n}$ is the TD error for quantile $n$ of critic $j$ for transition $i$ (computed as in \Cref{eq:critic_loss}), and $\epsilon$ is a small positive constant ensuring non-zero priority. Transitions are then sampled from the replay buffer $\mathcal{D}$ with probability $P(i)$ calculated as:
\begin{equation*}\label{eq:Pi}
P(i) = \frac{\delta_i^\omega}{\sum_{k \in \mathcal{D}} \delta_k^\omega},
\end{equation*}
where $\omega$ is a hyperparameter (see \Cref{tab:ml_params} in Appendix \ref{sec:detailed_params}) controlling the degree of prioritization ($\omega=0$ yields uniform sampling).
To correct for the bias introduced by this non-uniform sampling, importance sampling weights $w_i=1/(\mid\mathcal{D}\mid P(i)))^{\omega'}$ are applied during the loss calculations (where $\omega'$ typically anneals from an initial value to 1).
It should be noted that we use the same method for \gls{SACD}.

\subsection{Rewarding Off-Policy Methods}
Our training objective is to develop a deep scheduler that allocates scheduling candidates in both the frequency (\gls{RBG}) and spatial (MU-MIMO) domains in a proportionally fair manner.
One approach to achieve this is to maximize the long-term geometric mean of user throughput.
However, due to the necessity of making multiple \gls{RBG} decisions per MU-MIMO user layer at each TTI, we found it more effective to maximize the instantaneous increase in the \gls{PF} metric resulting from each scheduling decision.
Therefore, the reward for the $u$-th scheduling candidate on the $m$-th \gls{RBG} and $l$-th MU-MIMO user layer is defined as:
\begin{align} \label{eq:off-policy_reward}
\begin{split}
   r_{u,m,l} = \begin{cases}
   \dfrac{T_{u,m,l}}{R_u} - \dfrac{T_{u,m,l-1}}{R_u} & \text{for } l > 1 \\
   \dfrac{T_{u,m,l}}{R_u} & \text{otherwise}
   \end{cases}
\end{split},
\end{align}
where $T_{u,m,l}$ is the achievable data rate for that \gls{RBG} and user layer, and $R_u$ is the past average throughput of user $u$.

To prevent the model from learning to artificially minimize past average throughput to inflate rewards, we normalize the reward at each \gls{TTI}:
\begin{align} \label{eq:off-policy_reward_update}
\begin{split}
   r_{u,m,l} \leftarrow \begin{cases}
      \max\ \Biggl\{ \dfrac{r_{u,m,l}}{\max\limits_{u} r_{u,m,l}}, -1  \Biggl\} \quad \text{ for } \max\limits_{u} r_{u,m,l} > 0\\
      1 \qquad \qquad \; \text{ for } \max\limits_{u} r_{u,m,l} < 0 \text{ and } a = |A|\\
      -1 \qquad \quad \; \; \text{ otherwise}
   \end{cases}
\end{split},
\end{align}
This normalization clips the maximum reward to $1$ and the minimum to $-1$, ensuring a stable and balanced learning process.
We set the discount factor $\gamma=0$ intentionally to emphasize immediate \gls{PF} increments per \gls{RBG} decision.
Long‑term fairness is already enforced via the geomean KPI in evaluation.
Finally, the detailed training update steps for the \gls{DSACD} algorithm are presented in Algorithm \ref{alg:dsacd_v2}, Appendix \ref{sec:algorithms}.

 \section{Reference Scheduler Architecture}
\label{sec:reference_scheduler}

Building on the established practice of splitting radio scheduling into \gls{TDS} and \gls{FDS} as described in \cite{4556220}, our reference scheduler architecture adds an \gls{SDS} step to support \gls{MUMIMO}.
The \gls{TDS} step shortlists the candidate \glspl{UE} for scheduling based on \gls{QoS} requirements and, for example, the \gls{PF} metric. The \gls{FDS} step allocates frequency resources (optionally non-contiguous) to these candidates.
With \gls{MUMIMO}, the \gls{SDS} step pairs candidate \glspl{UE} on additional \gls{DL} \gls{MUMIMO} user layers for each \gls{RBG}.
We use two iterative heuristic \gls{SDS} methods similar to those in \cite{7022997}, modifying them to first allocate \gls{SUMIMO} resources with \gls{FDS} and then add spatial user layers until no suitable candidates remain or no additional \gls{MUMIMO} gain is expected with \gls{RZF}-based beamforming.
We describe both \gls{SDS} variants next:

\subsection{Baseline Spatial Domain Scheduler}
The baseline \gls{SDS} algorithm operates on a pre-sorted list of \gls{UE} candidates using the \gls{PF} metric from the \gls{TDS} phase.
It iterates through candidates in \gls{PF}-sorted order, selecting the first \gls{UE} that increases the total throughput of the \gls{RBG} when co-scheduled with \glspl{UE} already assigned to previous \gls{MUMIMO} user layers.
This approach can be computationally demanding for real-time environments, as it may need to iterate through all candidates to find one that improves sum throughput, posing a challenge for practical implementation.

\subsection{Proportional Fair Exhaustive Search per MU-MIMO Layer Spatial Domain Scheduler}
To explore further gains, we implemented a modified \gls{SDS} scheduler, \emph{\gls{PF} Greedy} \gls{SDS}. 
For each \gls{RBG} and \gls{MUMIMO} layer, this scheduler exhaustively searches for the candidate \gls{UE} that maximizes the \gls{PF} sum metric and increases the \gls{MUMIMO} sum throughput estimate for each \gls{RBG} and spatial layer.
This approach increases computational complexity, requiring recalculating beamforming weights and throughput estimations for every scheduled \gls{UE} and new candidate for each \gls{RBG} and spatial layer. 
While an exhaustive search over all MU-MIMO user layers could theoretically find the optimal combination, it is computationally infeasible even within our system-level simulations, let alone in a real-time environment.

\section{Computational Complexity Analysis}
\label{sec:complexity}

Table \ref{tab:ml_params} outlines the parameters affecting the computational complexity of both off- and on-policy methods.
When comparing the \gls{1L} and \gls{2L} architectures, it is important to clarify the definition of \emph{computational complexity} used herein.
While the \gls{2L} variant employs a neural network with fewer parameters per forward pass, its overall computational complexity per TTI is significantly higher due to the operational requirement of performing a separate inference for each \gls{RBG} and each \gls{MUMIMO} user layer.
This results in $N_{RBG} \times |L|$ forward passes per TTI for \gls{2L}, compared to only $|L|$ passes for \gls{1L}, which processes all \glspl{RBG}  for a given layer simultaneously.
Therefore, in this context, \emph{complexity} primarily refers to the total computational load and resulting inference latency per TTI, dominated by the number of required forward passes rather than solely the parameter count of the neural network used in a single pass.
This distinction is critical for real-time implementation constraints, as elaborated below and quantified in Table \ref{tab:execution_times}.

All input parameters are designed to avoid unnecessary per-\gls{TTI} calculations, as they can be pre-calculated based on \gls{CSI} reports. 
Additionally, \gls{UE} \gls{SUMIMO} precoder cross-correlations can be pre-calculated, with only the maximum value needed during the scheduling of additional \gls{MUMIMO} layers.
Consequently, execution times are primarily dependent on neural network forward passes.
Indicative forward pass times, measured using single-precision floating-point format neural networks on an Intel Xeon Gold 6330 CPU, are provided in Table \ref{tab:execution_times}. We cannot disclose results with real commercial \gls{BTS} hardware, but these equivalent measurements obtained from simulations using a CPU-optimized in-house C++ implementation, suggest that deep schedulers with \gls{2L} architectures may struggle to meet \gls{5G} inference latency requirements and \gls{TTI} limits.
In contrast, the \gls{1L} design offers a good balance between throughput performance and computational complexity, with scalability to massive MIMO.
We do not provide CPU times for the baselines because they depend on specific implementations and libraries, making direct comparisons unsuitable.

\begin{table}[H]
\centering
\caption{CPU times of deep scheduler forward passes with 18 \glspl{RBG} and 8 \gls{MUMIMO} \gls{UE} layers by using ReLU activation.}
\label{tab:execution_times}
\begin{tabular}{lccc}
\toprule 
\textbf{Model architecture} & \textbf{\gls{2L}} & \multicolumn{2}{c}{ \textbf{\gls{1L}} } \\ 
\midrule 
\textbf{Neurons per Hidden Layer} & \textbf{32} & \textbf{64} & \textbf{32}\\ 
\midrule 
CPU time per forward pass & 2 us & 5.6 us & 3.4 us \\ 
\midrule 
Sum CPU time per \gls{TTI} & 288 us & 44.8 us & 27.2 us \\ 
\bottomrule 
\end{tabular}
\end{table}

The execution times shown do not include state feature vector filling or other simulator or product-specific procedures.
However, due to the pre-calculated inputs, such preprocessing consumes a negligible amount of CPU time.
Forward pass times can be further reduced when models are used with CPUs supporting native half-precision floating-point operations.

Unlike heuristic baseline schedulers that require extensive throughput estimations with updated \gls{RZF} precoders and re-selected \glspl{MCS}, deep schedulers calculate \gls{RZF} precoders and select \gls{MCS} only once, after obtaining scheduling decisions from the actor networks.
This efficiency makes \gls{ML} models particularly appealing for scheduling tasks.

\newpage

\section{Pseudocode for Proposed RL Algorithms}
\label{sec:algorithms}
For clarity and reproducibility, the following algorithms detail the implementation of the training procedures for the \gls{1L}-\gls{PPO} (Algorithm \ref{alg:ppo_expert}) with expert guidance and \gls{DSACD} (Algorithm \ref{alg:dsacd_v2}) methods presented in Appendix \ref{sec:on-policy-framework} and Appendix \ref{sec:off-policy-framework}.

\begin{algorithm*}[!h]
\caption{1LDS-PPO with Expert Guidance Training}
\label{alg:ppo_expert}
\begin{algorithmic}[1]
\State Initialize actor network $\pi_\theta$, critic network $V_\phi$.
\State Initialize agent experience buffer $\mathcal{B}_\text{agent}$ (size $M=128$, for on-policy PPO updates).
\State Initialize expert demonstration buffer $\mathcal{D}_\text{expert}$ (size 4000 from \Cref{tab:ml_params} for \gls{JSD} loss).
\State Initialize expert policy $\pi_\text{expert}$ (e.g., PF scheduler).
\State Hyperparameters: PPO clip $\epsilon$, entropy coeff $\xi$, JSD weight $\lambda_\text{JSD}$, layers L.

\For{each environment TTI $t=1, 2, ...$}
    \State Initialize temporary storage for TTI trajectory $\mathcal{T}_\text{TTI}$.
    \For{layer $l=1$ to L}
        \State Observe state $s_{t,l}$ (Section III-B).
        \State Shuffle candidate UEs within $s_{t,l}$ if needed. \Comment{Perform Data Augmentation (optional, \Cref{sec:TrDataAugment})}
        \State Generate action probabilities $\mathbf{z}_{t,l} = \pi_\theta(s_{t,l})$. Sample action $\mathbf{a}_{t,l}$ (\Cref{sec:ActionSpace}).
        \State Generate expert action $\mathbf{a}_{\text{expert}, t,l} = \pi_\text{expert}(s_{t,l})$ (\Cref{sec:on-policy-framework}).
        \State Store $(s_{t,l}, \mathbf{a}_{t,l}, \text{log}\pi_\theta(\mathbf{a}_{t,l}|s_{t,l}), \mathbf{a}_{\text{expert}, t,l})$ in $\mathcal{T}_\text{TTI}$. \Comment{Store log prob for PPO}
        \State Resource allocation based on $\mathbf{a}_{t,l}$ happens here
    \EndFor
    \State \Comment{Environment step complete for TTI $t$. MCS selected, throughput $p_u$ realized.}
    \State \Comment{At TTI $t+1$, calculate rewards for TTI $t$:}
    \State Calculate $G$ using $p_u$ from TTI $t$ (\Cref{eqn:OptGoal}). Calculate $P$ (\Cref{eq:GM_throughput}).
    \For{each $(s_{t,l}, \mathbf{a}_{t,l}, \text{log}\pi_\text{old}, \mathbf{a}_{\text{expert}, t,l})$ in $\mathcal{T}_\text{TTI}$}
        \State Calculate reward $r_{t,l}$ using $P$ and $v_{m}$ (\Cref{eq:ppo_rwd}). \Comment{$v_m$ needs to be computed based on TTI $t$ outcome}
        \State Store $(s_{t,l}, \mathbf{a}_{t,l}, r_{t,l}, \text{log}\pi_\text{old})$ in agent buffer $\mathcal{B}_{agent}$. \Comment{Accumulates across TTIs}
        \State Store $(s_{t,l}, \mathbf{a}_{\text{expert}, t,l})$ in expert buffer $\mathcal{D}_\text{expert}$ (if buffer not full or using replacement).
    \EndFor

    \If{$|\mathcal{B}_\text{agent}|$ reaches update size $M$} \Comment{Perform Alternating Optimization Update}
        \State \Comment{PPO Update}
        \State Compute target returns $\hat{J}_i$ and advantages $\hat{A}_i$ for transitions in $\mathcal{B}_\text{agent}$ using \gls{GAE} \cite{gae}.
        \State Compute PPO loss $\mathcal{L}_\text{PPO}$ (\Cref{eq:ppo,eq:l_value,eq:l_policy,eq:l_entropy}) using transitions in $\mathcal{B}_{agent}$.
        \State Update $\theta, \phi \leftarrow \theta, \phi - \lambda_\text{PPO} \nabla_{\theta,\phi} \mathcal{L}_\text{PPO}$.
        \State Clear agent buffer $\mathcal{B}_\text{agent}$.

        \State \Comment{Expert Guidance Update}
        \State Sample mini-batch $\mathcal{B}_\text{expert} = \{(s_j, \mathbf{a}_{\text{expert},j})\}_{j=1..b'}$ from $\mathcal{D}_\text{expert}$. \Comment{$b'$ might differ from PPO batch}
        \State Compute JSD loss $\mathcal{L}_\text{JSD}$ (\Cref{eq:JSD_loss}) using $B_\text{expert}$ and current policy $\pi_\theta$.
        \State Update policy $\theta \leftarrow \theta - \lambda_\text{JSD} \nabla_\theta \mathcal{L}_\text{JSD}$.
    \EndIf
\EndFor
\end{algorithmic}
\end{algorithm*}

\begin{algorithm*}[!h]
\caption{DSACD Training Update}
\label{alg:dsacd_v2}
\begin{algorithmic}[1]
\State Initialize policy network $\pi_\theta$, critic networks $Q_{\phi_1}, Q_{\phi_2}$ (outputting $N$ quantiles per action), target networks $\overline{Q}_{\overline{\phi}_1}, \overline{Q}_{\overline{\phi}_2}$ with $\overline{\phi}_j \leftarrow \phi_j$.
\State Initialize entropy coefficient $\alpha$, replay buffer $\mathcal{D}$ (size from \Cref{tab:ml_params}), quantile levels $\tau_n = n/N$.
\State Hyperparameters: Mini-batch size $b$, PER $\omega$, target entropy $\beta$, learning rates $\lambda_Q, \lambda_\pi, \lambda_\alpha$, target update $\tau$.

\Procedure{UpdateNetworks}{}
    \If{$|\mathcal{D}| < b$} \Comment{\textbf{Wait for enough samples}}
        \State \textbf{return}
    \EndIf

    \State Sample mini-batch $B = \{(s_i, a_i, r_i, s'_i, \text{mask}_i, \text{mask}'_i)\}_{i=1...b}$ from $\mathcal{D}$ using sampling probabilities $P(i)$ 

    \State \Comment{\textbf{Critic Update}}
    \State Compute target quantile values $y_{\overline{\phi}, n}(r_i, s'_i)$ for each $n=1..N$ using Eq. \ref{eq:critic_target} \textit{(Requires target critics, policy $\pi_\theta$, $\alpha$)}.
    \For{$j=1, 2$}
        \State Compute TD errors $x_{\phi j, n, i} = q_{\phi_j, n}(s_i, a_i) - y_{\overline{\phi}, n}(r_i, s'_i)$.
        \State Compute critic loss $\mathcal{L}(\phi_j) = \frac{1}{bN} \sum_{i=1}^b \sum_{n=1}^N w_i \rho_{\tau_n}^k(x_{\phi j, n, i})$ using quantile Huber loss $\rho_{\tau_n}^k$ (Eq. \ref{eq:quantile_huber}).
        \State Update critic $\phi_j \leftarrow \phi_j - \lambda_Q \nabla_{\phi_j} \mathcal{L}(\phi_j)$.
    \EndFor

    \State \Comment{\textbf{Policy and Entropy Update}}
    \State Compute mean Q-values $Q_{\phi}(s_i, \cdot) \leftarrow \frac{1}{N}\sum_{n=1}^N \min_{j=1,2} q_{\phi_j, n}(s_i, \cdot)$ (Eq. \ref{eq:Q_target}).
    \State Compute policy loss $J_\pi(\theta) = \frac{1}{b} \sum_{i=1}^b w_i \mathbb{E}_{a \sim \pi_\theta(\cdot|s_i)}[\alpha \log(\pi_\theta(a|s_i)) - Q_{\phi}(s_i, a)]$ (Eq. \ref{eq:policy_objective}). Use action mask $\text{mask}_i$.
    \State Update policy $\theta \leftarrow \theta - \lambda_\pi \nabla_\theta J_\pi(\theta)$.

    \State Compute state-specific entropy target $\overline{H}(s_i) = -\beta \log (1 / |A(s_i)|)$ using Eq. \ref{eq:entropy_H} (where $|A(s_i)|$ from $\text{mask}_i$).
    \State Compute entropy loss $J(\alpha) = \frac{1}{b} \sum_{i=1}^b w_i \mathbb{E}_{a \sim \pi_\theta(\cdot|s_i)}[-\alpha (\log(\pi_\theta(a|s_i)) + \overline{H}(s_i))]$ (Eq. \ref{eq:enropy_objective}). Use action mask $\text{mask}_i$.
    \State Update entropy coefficient $\alpha \leftarrow \alpha - \lambda_\alpha \nabla_\alpha J(\alpha)$.

    \State \Comment{\textbf{Target Network Update}}
    \State Update target networks: $\overline{\phi}_j \leftarrow \tau \phi_j + (1-\tau) \overline{\phi}_j$ for $j=1, 2$.

    \State \Comment{\textbf{Priority Update}}
    \State Update sampling probabilities $p_i \propto (\frac{1}{2N}\sum_{j,n} |x_{\phi j, n, i}|)^\omega$ for used samples (Eq. \ref{eq:per_priority}).
    \State Update probabilities in $\mathcal{D}$ for sampled indices $i$.
\EndProcedure

\State \Comment{Main loop: Interact with env, store in $\mathcal{D}$, call UpdateNetworks periodically}
\end{algorithmic}
\end{algorithm*}


 \section{Detailed Simulation Parameterizations}
 \label{sec:detailed_params}

\begin{table}[H]
\centering
\caption{System parameters}
\label{tab:sys_params} 
\begin{tabular}{ll} 
\toprule
\multicolumn{2}{c}{\bfseries Evaluation Parameters}  \\
\bfseries Parameter & \bfseries Value \\
\midrule
Deployment & 3GPP Dense Urban Macro (see Table 6.1.2-1 in \cite{3gpptr38913}) \\
Number of cells & 21\\
Inter-site distance & 200 m \\
gNB height & 25 m\\
gNB Tx power & 44 dBm \\
Number of \glspl{UE} & 210 uniformly distributed \\
Random drops & 10 \\
\midrule
Bandwidth &  100 MHz (273 RBs, $N_{\text{RBG}} = 18$)\\ 
Carrier frequency & 4 GHz\\ 
Subcarrier spacing & 30 kHz \\
\midrule
MIMO scheme & \gls{MUMIMO} with regularized Zero Forcing\\
Modulation & QPSK to 256QAM\\
Link adaptation & 10 \% \gls{BLER} target \\
gNB antenna panel & 12x8 and 2 polarizations\\
Max \gls{UE} layers & $|L| = 8$ \\
Max \gls{UE} rank & 2 \\
Max MIMO layers & 16 (i.e. $|L| \times \text{max rank}$)\\
\midrule
\gls{CSI} & \gls{RI} and sub-band \gls{CQI} \& \gls{PMI} \\
Max scheduled \glspl{UE} & $|U| = 10$ \\
\gls{TD} scheduler & \gls{PF} \\
\midrule
UE speed & 3 km/h\\
UE height & 1.5 m\\
UE receiver & \gls{MRC}\\
UE antenna & 2 dual-polarized Rx antennas\\
\midrule
Traffic model & 100\% Full Buffer, or 100\% FTP Model 3 \cite{3gpptr36889}\\
{} & (FTP3 with 0.5 MB packets, 20 packets/s)\\
\midrule
\multicolumn{2}{c}{\bfseries Parameter Modifications for Training}  \\
\bfseries Parameter & \bfseries Value \\
\midrule
Number of \glspl{UE} & 420 uniformly distributed \\
Random drops & 1 \\
Traffic model & 50\% of \glspl{UE} Full Buffer and \\
{} & 50\% of \glspl{UE} FTP Model 3 \cite{3gpptr36889}\\
{} &  (FTP3 with  1.5 kB packets, 500 packets/s)\\
Bandwidth &  6.6 MHz (18 RBs, 18 RBGs)\\
Max \gls{UE} layers & $|L| = 4$ \\
\bottomrule 
\end{tabular}
\end{table}

\begin{table}[H]
\centering
\caption{Training hyperparameters}
\label{tab:ml_params} 
\begin{tabular}{lccc} 
\toprule 
\textbf{Parameter} & \multicolumn{2}{c}{\textbf{Off-Policy}} & \textbf{PPO} \\
\midrule 
Discount factor & \multicolumn{2}{c}{0} & 0.95 \\
Epochs per sample & \multicolumn{2}{c}{1} &  1 \\
Mini batch size& \multicolumn{2}{c}{32} &  128 \\
Actor learning rate & \multicolumn{2}{c}{0.0001} &  0.0001 \\ 
Critic learning rate & \multicolumn{2}{c}{0.0001} &  0.0002 \\ 
Hidden layers & \multicolumn{2}{c}{2} & 2 \\
Hidden layer size & \multicolumn{2}{c}{32} & 32 \\
Activation function & \multicolumn{2}{c}{ReLU} & ReLU \\
Optimizer &  \multicolumn{2}{c}{Adam \cite{Adam}} & Adam \cite{Adam} \\
Replay buffer size &  \multicolumn{2}{c}{$N_{\text{gNB}} \times 1000$} & 4000 \\
Prioritized replay $\omega$ & \multicolumn{2}{c}{0.5}  & N.A. \\
Target smoothing coef. & \multicolumn{2}{c}{0.001} & N.A. \\
Reward scaling factor $k$ & \multicolumn{2}{c}{N.A.} & 0.2 \\
Clipping parameter $\epsilon$ in PPO & \multicolumn{2}{c}{N.A.} & 0.2 \\
\midrule 
{} & \textbf{\gls{SACD}} & \textbf{\gls{DSACD}} & \textbf{\gls{PPO}} \\
\midrule 
Critic quantiles $N$ & N.A. & 16 & N.A. \\
\bottomrule 
\end{tabular}
\end{table}

\begin{table}[H]
\centering
\caption{Architecture hyperparameters}
\label{tab:arch_params} 
\begin{tabular}{lcc} 
\toprule 
{} & \textbf{\gls{2L}} & \textbf{\gls{1L}} \\
\midrule 
Input size & $|U| \times 8 + 1$ & $|U| \times (5 + 2N_{\text{RBG}})$ \\
Output size & $|U|+1$ & $N_{\text{RBG}} \times (|U|+1)$ \\
Inferences per \gls{TTI} & $|L|\times N_{\text{RBG}}$ & $|L|$ \\
\gls{SAC} target entropy $\beta$ & 0.4 & 0.999 \\
\bottomrule 
\end{tabular}
\end{table}

\end{document}